\documentclass[11pt]{article}
\usepackage{cite}
\usepackage{amssymb,amsmath,amsthm,mathrsfs}

\usepackage[english]{babel}
\usepackage{graphicx}
\usepackage{dcolumn}
\usepackage{bm}
\usepackage{color}
\usepackage{dsfont}
\usepackage{subfigure}
\usepackage[ansinew]{inputenc}
\usepackage{amsfonts}
\usepackage{amsmath} 
\def\<{\langle}
\def\>{\rangle}

\def\ket#1{| #1 \rangle}

\def\e{\mathrm{e}}

\def\e{e}
\def\i{i}

\newtheorem{Remark}{Remark}

\begin{document}

\title{Open quantum dynamics without Complete Positivity: a criticism}

\author{Fabio Benatti \\
Dipartimento di Fisica, Università di Trieste, Trieste, Italy \\
INFN, Sezione di Trieste, Trieste, Italy\\[1ex]
email: benatti@ts.infn.it \\ \\
Dariusz Chru\'sci\'nski\\
Institute of Physics, Nicolaus Copernicus University, Toru\'n, Poland
\\[1 ex]
e-mail: darch@fizyka.umk.pl
\\ \\
Saverio Pascazio \\
Dipartimento di Fisica, Universit\`{a} di Bari, Bari, Italy \\
INFN, Sezione di Bari, Bari, Italy\\[1ex]
e-mail: saverio.pascazio@ba.infn.it}
\author{Fabio Benatti$\hspace{0.4mm}^{1,2}$\thanks{Email: benatti@ts.infn.it}\;, \hspace{1.2mm}
\hspace{1.2mm}
Dariusz Chrusci\'nski$\hspace{0.4mm}^3$\thanks{Email:darch@fizyka.umk.pl}\,;\hspace{1.2mm}
\hspace{1.2mm}
Saverio Pascazio$\hspace{0.4mm}^{4,5}$\thanks{Email: saverio.pascazio@uniba.it}}

\date{
{\normalsize
$^1$\textit{Department of Physics, University of Trieste, I-34151, Trieste, Italy}\\	
\vspace{2mm}
$^2$\textit{Istituto Nazionale di Fisica Nucleare, Sezione di Trieste, I-34151,\\
Trieste, Italy}\\[1ex]e-mail:benatti@ts.infn.it}\\
\vspace{2mm}
$^3$\textit{Institute of Physics, Nicolaos Copernicus University, Torun, Poland}\\[1ex]e-mail: darch@fizyka.umk.pl\\
\vspace{2mm}
$^4$\textit{Department of Physics, University of  Bari, I-70126 Bari, Italy}\\
\vspace{2mm}
$^5$\textit{Istituto Nazionale di Fisica Nucleare, Sezione di Bari, I-70126,\\
Bari, Italy}\\[1ex]
e-mail: saverio.pascazio@uniba.it}

\maketitle
	
\begin{abstract}  The requirement of complete positivity is very often regarded as a fundamental consistency condition for the description of open quantum dynamics. We critically examine this requirement and discuss both its physical motivations and its limitations. We analyze proposals based on restricting the domain of non-completely positive maps to subsets of compatible initial states. Using isotropic states as a concrete example, we show that such domain restrictions become increasingly severe with growing system dimension, revealing an intrinsic weakness of the compatibility-based approach. 
\end{abstract}

\tableofcontents

%%================================================================================================================================================

\section{Introduction and motivations}
\label{intro}

\noindent
In Physics, when one formulates an evolution law in terms of a differential equation, one implicitly assumes that the law is valid for any initial condition. This is Laplace's legacy. A good example is the fundamental equation of mechanics
$F = ma = m \ddot x$.
Physics students are taught how to solve this equation for every initial condition
$x(t=0) = x_0 , \quad \dot x (t=0) = v_0$, where $x_0$ and $v_0$ are the initial position and velocity, respectively. This is called a Cauchy problem and one requires that the solution exists and is unique $\forall \; x_0, v_0$. 

When the solution does not exist or is not unique one faces a serious problem. An example is 
$\ddot x  = 6 x^{1/3}$, with $x(t=0) = 0 , \quad 
\dot x (t=0) = 0$,
that admits the two solutions $x_1(t) = 0$ and $x_2 (t) = t^3$.
One would be tempted to assume that a potential of the type $V \sim x^{4/3}$ does not exist in Nature (one would say that it is ``unphysical"), but unfortunately such a potential can be conceived, and this has generated controversy \cite{norton}.

There is a mathematical way out of this conundrum: the function $x^{1/3}$ is not Lipschitz near $x=0$ and hence the uniqueness theorem does not apply; therefore one assumes that physical forces and potential must satisfy the Lipschitz condition and removes ambiguity-generating initial conditions from the outset. This does not solve the physical problem though. A more physical way out would be to argue that an equation like the afore-mentioned one does not describe the behavior of a point particle with infinite accuracy, more so close to the origin, where the effects of microscopic details and friction (and even symmetry breaking) become crucial. If duly taken into account, the microscopic features of the environment would yield a unique solution.

Why do physicists not like the above example? Probably because one does not like the idea that the mechanical particle can 
``decide" its own trajectory, in a world, that of classical mechanics, that is deterministic.
This attitude goes very far: a Cauchy formulation, for a physical law describing time evolutions, is taken to be valid even for partial differential equations, say the wave equation. Also in this case, one wants that the solution exists and is unique. 
In full generality, abiding by such a request leaves two possible scenarios: i) either one should be open  to restricting the possible initial conditions to those that allow for the existence and uniqueness of trajectories, or, ii) one should modify the dynamics in order to allow for all possible initial conditions. 

%\rev{In fact, what we are saying here is not entirely correct (and is not the end of the story): one should rather say: ``ii) one should modify the dynamics in order to allow for all possible initial conditions of the preceding (unmodified, and less accurate) dynamics." See the argument of two paragraphs above. I find it curious that this classical problem is solved by introducing a better description of the environment. I leave it up to you, pls decide what to write.) }

The aim of this brief introduction is to show the similarity of such a dilemma with the request  of complete positivity (CP) in open quantum dynamics, specifically in the case of quantum dynamical semigroups. There, restricting the dynamics to those enacted by the generators of the so-called Gorini-Kossakowski-Sudarshan-Lindblad (GKLS) form~\cite{GKS,Lindblad,CP2017} goes along the lines of the second scenario ii) depicted above, namely paying the price of selecting as physically consistent only specific dynamical maps in order to use them to describe the time-evolution of any initial quantum state. Note that CP plays a crucial role in this context, because the system undergoing the dynamics could be entangled to parts of its environment, and a positive (non-CP) dynamics would not guarantee the positivity of the density matrix. See the following Section for additional considerations on this issue.

On the other hand, a quantum counterpart of the classical restriction of the initial conditions , scenario i) above, was proposed by a number of authors~\cite{Buzek,Shaji1,Shaji2,Carteret,Lidar} and by the advocates of the so-called "slippage of initial conditions"~\cite{HartmannStrunz,Suarez,Gaspard}.
Though not equivalent, all these proposals share a common spirit in that they restrict the initial states to those which are compatible with the dynamics, be it only positive (namely, not completely positive) or not even positive. {Here, states compatible with a given dynamics are those states that remain states, that is, density matrices with positive spectrum adding to $1$ for all times.}

We shall focus upon the first scenario above; to start with, we briefly survey the issue of complete positivity as it comes to the fore when dealing with open quantum systems.

\section{Notation and additional considerations}
\label{notas}

The evolution of a closed (isolated) quantum system is described by the so-called von Neumann equation for the density matrix $\rho_t$, describing the statistics of its  state at time $t$, starting from an initial condition $\rho$ at time $t=0$:
\begin{equation}
\label{vn}
\dot \rho_t \;=\; -\i [H,\rho_t] \quad \longleftrightarrow \quad \rho_t \;=\; U_t \rho_0 U^\dagger_t\,;\quad U_t =\exp{(-i\,H\, t)}\ .
\end{equation}
If the quantum system is ``open", namely not isolated, and immersed in an environment with which it interacts, in accordance with the request of Complete Positivity (see below), the one-parameter group of unitary maps
\begin{equation}
\label{unitaries}
\mathcal{U}_t:\rho\mapsto \mathcal{U}_t[\rho]=\rho_t
\end{equation}
is replaced by a one-parameter family of maps
\begin{equation}
\label{superoperator}
\rho_t \;=\; \Lambda_t[\rho] \;=\; \sum_\alpha K_\alpha(t)\,  \rho\, K_\alpha^\dagger(t) ,\quad \sum_\alpha K_\alpha^\dagger(t)  K_\alpha(t) = \mathds{I}\ .
\end{equation}
These maps are said to be in Kraus-Stinespring form~\cite{CP,KRAUS}.
Notice that if the summation consists of a single addendum $K(t) = U_t = \e^{-\i H t}$, (\ref{superoperator}) reduces to (\ref{vn}). Notice also that (\ref{superoperator}) is not a differential equation, but rather a ``snapshot" of the quantum state at a particular time $t$.

The above specific form  guarantees not only that $\Lambda_t$ is positive, but also that it is Completely Positive.
Namely, a map of the form $\Lambda_t$ as in~\eqref{superoperator} not only preserves the positivity of all states of an open system $S$ in the course of time, as it should do in order to allow for the association of its eigenvalues with probabilities.  What is more to it is that, when tensorized with the identical map ${\rm id}_n$, $\Lambda_t\otimes{\rm id}_n$ also 
preserves the positivity of all states of the compound system consisting of $S$ and an arbitrary, dynamically inert and dynamically decoupled, auxiliary $n$-level system $S_n$. 

Notice that, though not evolving and not interacting with $S$ (this is expressed by the factorization 
of the dynamical map of $S+S_n$ into the given one of $S$ and the trivial one of $S_n$), yet this auxiliary system can be statistically correlated to $S$.
Indeed, the reason behind the request of Complete Positivity, namely behind the fact that $\Lambda_t$ must be  of the Kraus-Stinespring form~\eqref{superoperator}, is that, otherwise, there certainly exists~\cite{Choi1,Choi2} an entangled state $\rho^{ent}_{S+S_n}$ of the compound system  $S+S_n$ some of whose eigenvalues becomes negative at some time, precluding the interpretation of $\Lambda_t\otimes{\rm id}_n\Big[\rho^{ent}_{S+S_n}\Big]$ as a physically consistent state.

Two aspects are worth stressing: the first one is that that not all entangled states of $S+S_n$ are incompatible with the compound dynamics 
$\Lambda_t\otimes{\rm id}_n$ when $\Lambda_t$ is not completely positive, as well as not all states of $S$ will be incompatible  with a map $\Lambda_t$ which is not even positive.
One might then argue, as done in~\cite{Buzek,Shaji1,Shaji2,HartmannStrunz,Suarez, Gaspard}, that, by restricting the dynamics to compatible states, namely to those initial states whose spectrum remains positive under $\Lambda_t\otimes{\rm id}_n$ or $\Lambda_t$, one might do without the Complete Positivity or even the Positivity of $\Lambda_t$.

The second aspect concerns the fact that completely positive $\Lambda_t$ as in~\eqref{superoperator} are very much constrained.
In the Markovian approximation, when the maps $\Lambda_t$ compose as a semi-group,  the form~\eqref{superoperator} emerges from the so-called rotating way approximation as in the weak-coupling limit prescription~\cite{DAVIES,DumckeSpohn,Alicki}. Otherwise, as for the Redfield equations~\cite{Redfield}, the ensuing maps $\Lambda_t$ are, in general,  not even positive.

As we shall see below, the Kraus-Stinespring form of the dynamics has dramatic consequences on the physics of an open quantum system $S$. 
{Foremost is} the appearance of a hierarchy in its decay properties~\cite{Alicki}.
As they arise from the request of complete positivity, these consequences are only due to the request that the dynamics be compatible with the possible non-local correlations between $S$ and any arbitrary auxiliary, inert, $n$-level systems $S_n$.
Due to this fact, instead of seeking dynamics compatible with such a hardly concretely controllable possibility, one started looking for physical states compatible with dynamics free from the constraints imposed by the Kraus-Stinespring form~\eqref{superoperator}. Complete positivity started then to be
challenged as an unnecessary mathematical simplification, not entirely  physically justifiable.

However, in the following we argue that, though one might not be afraid of the possible lack of Complete Positivity, yet the latter cannot be easily dispensed with.
Indeed, after a short survey of the pros and cons of Complete Positivity, and how one can dismiss it by restricting to subsets of compatible states, 
we show in a concrete and simple example that these subsets may become thinner and thinner with increasing dimension.
Certainly, this argument does not solve the issue, but emphasizes that 1) Complete Positivity has to do with entanglement and 2) the existence of 
non-local  quantum correlations in high dimension may make non completely positive dynamics conflict with the existence of mesoscopic and macroscopic entanglement.

\section{Complete positivity: Pros, Cons and an Escape from Negative Probabilities}

We shall consider finite, $d$-level quantum systems $S$, described by a Hilbert space $\mathcal{H}=\mathbb{C}^d$ and denote by $\mathcal{S}(S)$ the convex set of its states $\rho$ ($d\times d$ density matrices), by $\mathcal{M}_d$ the $d\times d$ matrix algebra of its observables $X$ and by $\Lambda$  the linear maps from $\mathcal{M}$ onto itself.

The states $\rho\in\mathcal{S}$ are positive operators of trace $1$, whose eigenvalues are then positive and sum up to $1$; as probabilities of finding the system $S$ in the corresponding eigenstates, they are at the root of the statistical interpretation of quantum mechanics.

Under a linear map $\Lambda$ on $\mathcal{M}$ that describes a transformation of the observables of $S$, its states change by duality according to
$\rho\mapsto\Lambda^\#[\rho]$ where $\Lambda^\#$ is also linear map on $\mathcal{S}(S)$ such that ${\rm Tr}(\rho\Lambda[X])={\rm Tr}(\Lambda^\#[\rho]\,X)$.
It follows that, in order for $\Lambda$ to describe an actual physical transformation,  $\Lambda^\#$ must map states into states, thus $\mathcal{S}(S)$ into itself, preserving their positivity.
Such maps are called Positive~\footnote{They share this property with the maps $\Lambda$ of which they are dual.}: non-positive maps then make negative eigenvalues appear in the spectrum of $\Lambda^\#[\rho]$ for at least one state $\rho\in\mathcal{S}(S)$ spoiling its probabilistic interpretation, when transformed.

A Positive linear map $\Lambda$ has the stronger property, known as Complete Positivity, if, for all $n\geq 1$, $\Lambda\otimes {\rm id}_n$ is Positive 
as a linear  map on $\mathcal{M}\otimes M_n$, where ${\rm id}_n$ denotes the identity operation on the $n\times n$  matrix algebra $M_n$.
Namely, $\Lambda$ is completely positive if, when lifted to act trivially on any ancillary finite-level system $S_n$ statistically correlated to the system $S$, its dual $\Lambda^\#\otimes{\rm id}_n$  does not alter the positivity of the states of the compound system $S+S_n$ and thus maps the space of states 
$\mathcal{S}(S+S_n)$ into itself.

As already outlined in the Introduction, the finite level system $S_n$ appended to $S$ does not evolve either by itself or by  dynamically coupling it to $S$: the only non-trivial fact about its presence is that the two may share non-local correlations. 
The issue at stake here, and challenged by~\cite{Buzek,Shaji1,Shaji2,HartmannStrunz,Suarez,Gaspard}, is why, exactly because of such a possibility,  one should require the Complete Positivity of $\Lambda$ in order to accept $\Lambda$ as a physically consistent transformation.

\subsection{Pros}

The fundamental reason for such a request is the existence of entanglement between $S$ and $S_n$~\cite{HHHH}: 
indeed, if all compound states of $S+S_n$ were separable, namely of the form
\begin{equation}
\label{add1}
\rho_{S+S_n}=\sum_j\lambda_j\rho_S^j\otimes\rho_{S_n}^j\ ,\quad 0\leq\lambda_j\leq 1\ ,\quad \sum_j\lambda_j=1\ ,
\end{equation}
with $\rho_S^j$ and $\rho_{S_n}^j$ states of $S$ and $S_n$, respectively; then, the positivity of $\Lambda^\#$ would be sufficient to guarantee  that its 
so-called lifting $\Lambda^\#\otimes{\rm id}_n$ also maps states of $S+S_n$ into states. Indeed,
\begin{equation}
\label{add2}
\Lambda^\#\otimes{\rm id}_n[\rho_{S+S_n}]=\sum_j\lambda_j\Lambda^\#[\rho_S^j]\otimes\rho_{S_n}^j
\end{equation}
is positive for such are the operators $\Lambda^\#[\rho_S^j]$ in the above convex decomposition.

That entanglement is at the root of the physical justification of the need of complete  positivity is epitomized by taking $\Lambda$ as the transposition over a $d$-level quantum system; then, the lifted action $T\otimes{\rm id}_d$ on $S_d+S_d$, known as partial transposition, is such that the projection $P^{(d)}_{sym}$ onto the totally symmetric entangled vector state
\begin{equation}
\label{symvec}
\vert\Psi_{sym}^{(d)}\rangle=\frac{1}{\sqrt{d}}\sum_{j=1}^d\vert j\rangle\otimes\vert j\rangle
\end{equation}
is mapped into an operator $T\otimes{\rm id}_d[P^{(d)}_{sym}]$ proportional to the flip operator
\begin{equation}
\label{add3}
V=\sum_{j,k=1}^d\vert k\rangle\langle j\vert\otimes\vert j\rangle\langle k\vert
\end{equation}
that has a $d(d-1)/2$ degenerate negative eigenvalues $-1$ and cannot therefore belong to $\mathcal{S}(S+S_n)$.

We can summarize the pros of complete positivity as follows:
\begin{enumerate}
\item
linear maps $\Lambda$ describe physical transformations of $S$ if they preserve the positivity of all  states in $\mathcal{S}(S)$;
\item
since $S$ cannot be prevented from being statistically correlated to an arbitrary $n$-level system $S_n$, which is otherwise inert,  positivity must also be true of $\Lambda\otimes{\rm id}_n$ as describing a physical transformation of all states in  $\mathcal{S}(S+S_n)$, for all $n\in\mathbb{N}$.
\end{enumerate}

In particular, the Choi theorem \cite{Choi2} states that a maps $\Lambda:M_d\mapsto M_d$ is completely positive if and only if 
its Choi matrix 
\begin{equation}
\label{Choi}
C_{\Lambda}:=\Lambda\otimes{\rm id}_d[P^{(d)}_{sym}]\geq 0\ .
\end{equation}
This means that if $\Lambda$ is not completely positive, certainly the entangled state of $S_d+S_d$ represented by the projector $P^{(d)}_{sym}$ onto 
the totally symmetric state vector in~\eqref{symvec}
is mapped out of the space of states $\mathcal{S}(S_d+S_d)$ by the lifted map $\Lambda\otimes{\rm id}_d$.

\subsection{Cons}

Complete positivity has a strong structural consequence; indeed, unlike simply positive maps, whose generic structure is still unknown, all completely positive linear maps must be of the so-called Kraus-Stinespring form
\begin{equation}
\label{KS}
\Lambda[x]=\sum_\alpha L^\dag _\alpha\,x\, L_\alpha\ ,
\end{equation}
where $L_\alpha$ are operators on the system Hilbert space such that the sum $\sum_\alpha L^\dag_\alpha\,L_\alpha$ exists in a suitable topology.

The specific form of completely positive maps is rather constraining with severe physical consequences; in the case of continuous semigroups $\Lambda_t=\exp(t\mathbb{L}): M_d\mapsto M_d$, CP is equivalent to the fact that, in the GKLS generator
\begin{equation}
\label{GKSL}
\mathbb{L}[x]=i[H\,,\,x]+\sum_{\alpha,\beta=1}^{d^2-1}C_{\alpha\beta}\,\left(F_\alpha^\dag\,x\,F_\beta-\frac{1}{2}\left\{F_\alpha^\dag F_\beta\,,\,x\right\}\right)\ ,
\end{equation}
where $F_\alpha\in M_d$ are traceless matrices such that ${\rm Tr}(F^\dag_\alpha\,F_\beta)=\delta_{\alpha\beta}$, the $(d^2-1)\times(d^2-1)$ Kossakowski 
matrix of coefficients $C=[C_{\alpha\beta}]$ must be positive semi-definite.

The adversaries of Complete Positivity argue that its resulting constraints are more a physical inconvenience than an appreciable property. Indeed, the decay-time hierarchy is enforced only on the basis of the hypothetical  entanglement of an open system with an uncontrollable and inert ancillary system;  such a mere possibility may nevertheless having a strong impact on,  for instance, the way the open system tends to equilibrium.

As a concrete example, consider an open qubit ($d=2$) subject to the purely dissipative dynamics generated by~\eqref{GKSL} with $H=0$, $\sqrt{2}\,F_{1,2,3}$ the Pauli matrices $\boldsymbol{\sigma}=(\sigma_1,\sigma_2,\sigma_3)$,
$C=\hbox{diag}(1,1,a)$ and $a\geq 0$. The master equation $\partial_t\rho_t=\mathbb{L}[\rho_t]$ is easily solved in the Bloch representation yielding
$$
\rho=\frac{1+\bf{r}\cdot\boldsymbol{\sigma}}{2}\mapsto \rho_t=\Lambda_t[\rho]=\frac{1+{\bf r}_t\cdot\boldsymbol{\sigma}}{2}\ ,
$$
with Bloch vector
$$
{\bf r}_t=\left({\rm e}^{-(1+a)t}\,r_1,{\rm e}^{-(1+a)t}\,r_2,{\rm e}^{-2t}r_3\right)\ .
$$
Then, the condition for Positivity $\|{\bf r}_t\|\leq 1$ is satisfied by $a\geq -1$, while CP requires $a\geq 0$: the elimination of the parameters $-1\leq a<0$ from those with a physical interpretation is only due to the possibility that the qubit of interest and evolving according to $\Lambda_t$ be initially entangled with another, dynamically insensitive, qubit, so that the two qubits together evolve according to $\Lambda_t\otimes{\rm id}_2$.

We have thus far gathered the following information: 1) complete positivity is enforced by the necessity of forbidding the appearance of negative probabilities when considering the possible entanglement of a given system $S$ with a generic auxiliary finite-level system $S_n$; 2) the auxiliary system is  so generic, possibly distant and  dynamically inert, to be practically uncontrollable; 3) in spite of this,  the presence of in line of principle uncontrollable non-local correlations of $S$ with $S_n$  must nonetheless enforce appreciable physical consequences; 4) for instance, they can  forbid certain types of decay that 
may conflict with a priori expectations about the way an open quantum system reaches asymptotic equilibrium.

Based on the above points, complete positivity is often refused as an unnecessary mathematical simplification, whose many constraints may lose sight of the actual open quantum dynamics. 

\subsection{A Way-out}

Suppose one would like to physically accept a non-CP map $\Lambda$, for it more satisfactorily describes the behavior of certain states of an open quantum system. How can then one cope with the negative probabilities that pop up when, for instance, one considers the totally symmetric projector $P^{(d)}_{sym}$ and let it evolve under $\Lambda\otimes{\rm id}_d$?

Certainly not all entangled states of $S_d+S_d$ behave as badly as $P^{(d)}_{sym}$ under $\Lambda\otimes{\rm id}_d$: 
borrowing the terminology from~\cite{Shaji1}, let us call $\Lambda$-incompatible all those entangled states that are mapped out of $\mathcal{S}(S_d+S_d)$ by $\Lambda\otimes{\rm id}_d$,  and $\Lambda$-compatible the complementary set in
$\mathcal{S}(S_d+S_d)$.
Then, a way out of the negative probability issue is to restrict the initial states to only those that are $\Lambda$-compatible.

As a concrete instance, suppose one would like to salvage transposition $T$ as a physically consistent map on $M_d$; in order to avoid $T\otimes{\rm id}_d$ 
to push initial states out of the space of states of $S+S_d$, one restricts it to act only on the subset of states of  $\mathcal{S}(S_d+S_d)$ that are Positive under Partial Transposition (PPT).
Of course, one should eliminate from the domain of $T\otimes{\rm id}_d$ all those (NPT) states that do Not remain Positive under Partial Transposition.

This clearly indicates that accepting non-CP maps $\Lambda$ by restricting their domain of application might exclude large sets of physically interesting states.
It is on the basis of such considerations that, in the following, we point to a weakness in the compatibility domain approach~\cite{Buzek,Shaji1,Shaji2} to 
circumventing the need of Complete Positivity.

\section{A possible obstruction to $\Lambda$-compatibility: increasing dimension}

In order to make manifest the afore-mentioned weakness in the $\Lambda$-compatible subset approach, we consider the class of isotropic qudit states introduced in~\cite{Horodecky}:
\begin{equation}
\label{Horostates}
\rho_F=\frac{1-F}{d^2-1}\mathds{1}_{d^2}\,+\,\frac{d^2\,F-1}{d^2-1}\,P^{(d)}_{sym}\ ,
\end{equation}
where  %\rev{isn't $1/d^2 \leq F$?} 
$P^{(d)}_{sym}$ projects onto the totally symmetric two-qudit state
\begin{equation}
\ket{\Psi^{(d)}_{sym}}=\frac{1}{\sqrt{d}}\sum_{j=1}^d\ket{j}\otimes\ket{j}\ ,
\label{TotSym}
\end{equation}
given a fixed orthonormal basis $\{\ket{j}\}\in\mathbb{C}^d$. Positivity of $\rho_F$ amounts to $0\leq F\leq 1$: indeed, its eigenvalues are
$\displaystyle \frac{1-F}{d^2-1}$ and $\displaystyle \frac{1-F}{d^2-1}+\frac{d^2\,F-1}{d^2-1}=F$.
Furthermore, one knows that the partial transposition is an exhaustive witness for such states and, 
consequently, that  the isotropic states are separable for $0\leq F\leq 1/d$ and entangled for $1/d<F\leq 1$.

\begin{Remark}
\label{rem1}
Already at this stage, notice that no entangled isotropic states can be salvaged from being mapped out of the space of states by the 
partial transposition $T\otimes{\rm id}$. 
From the point of view in~\cite{Shaji2}, none of them  would then be $T$-compatible.
Moreover, taking the length $(d-1)/d$ of the interval of the parameter $F$ corresponding to entangled $\rho_F$ as a measure of their relative volume,
one sees that, the larger the dimension $d$, the closer to $1$ is the volume of isotropic states spoiled by partial transposition.
\end{Remark}

We shall now make use of the following one parameter family of positive and unital linear maps from $M_d(\mathbb{C})$ into $M_d(\mathbb{C})$,
\begin{equation}
\label{PCPmaps}
\Lambda_\mu = \mu\frac{1}{d}\, {\rm Tr}\,+\,(1-\mu)\, {\rm T}\ ,\quad \Lambda_\mu[\mathbb{I}]=\mathbb{I}\ ,
\end{equation}
given by convex combinations with $0\leq\mu\leq 1$ of  the (positive) normalized trace-map $\displaystyle\,x\,\mapsto\frac{1}{d}{\rm Tr}[x]$ and of the (positive) transposition  ${\rm T}$.
This map is self-dual, that is $\Lambda^\#_\mu=\Lambda_\mu$ and its Choi matrix is given by
\begin{equation}
\label{Choi-mu}
C_\mu=\Lambda_\mu\otimes{\rm id}[P^{(d)}_{sym}]=\frac{\mu}{d^2}\,\mathds{1}_{d^2}+\frac{1-\mu}{d}\,V\ ,
\end{equation}
where $V\,\ket{\psi\otimes\phi}=\ket{\phi\otimes\psi}$ is the flip operator with $d(d+1)/2$ eigenvaules $+1$ and $d(d-1)/2$ eigenvalues $-1$.
The spectrum of $C_\mu$ thus consists of eigenvalues
\begin{equation}
\label{eigCmu}
\left\{\begin{matrix}
\displaystyle
\frac{\mu}{d^2}+\frac{1-\mu}{d}=\frac{d-\mu(d-1)}{d^2}&\hbox{with degeneracy}\ \frac{d(d+1)}{2}\cr
\cr
\displaystyle 
\frac{\mu}{d^2}-\frac{1-\mu}{d}=\frac{\mu(1+d)-d}{d^2}&\hbox{with degeneracy}\  \frac{d(d-1)}{2}\end{matrix}
\right.\ .
\end{equation}
From the Choi  theorem \cite{Choi2} the Positive maps $\Lambda_\mu$ cannot be Completely Positive in the presence of negative eigenvalues of $C_\mu$; thus, for 
\begin{equation}
\label{CPvsP}
0\leq\mu \leq \frac{d}{1+d}\ ,\quad \frac{d}{d-1}\leq \mu\leq 1\ .
\end{equation}
For these values of $\mu$, we are now interested in the volume of the entangled isotropic states that are compatible with Positive but not Completely Positive $\Lambda_\mu$; such volume can be computed by evaluating the positivity of the spectrum of
\begin{eqnarray*}
\label{comp-isotr-mu1}
\Lambda_\mu\otimes{\rm id}_d[\rho_F]&=&\frac{1-F}{d^2-1}\,\mathds{1}_{d^2}\,+\,\frac{d^2\,F-1}{d^2-1}\left(\frac{\mu}{d^2}\,\mathds{1}_{d^2}\,+\,\frac{1-\mu}{d}\,V\right)\\
\label{comp-isotr-mu2}
&=&
\frac{d^2(1-F)+\mu(d^2F-1)}{{d^2(d^2-1)}}+\frac{d(d^2F-1)(1-\mu)}{{d^2(d^2-1)}}\,V\ .
\end{eqnarray*}
It consists of the two degenerate eigenvalues 
\begin{equation}
\label{eigenvs}
E_\pm=\frac{d\pm d^2F+\mu(d^2F\pm1)}{d^2(d\pm1)}\ .
\end{equation}
While $E_+\geq 0$, the positivity of $E_-$ requires
\begin{equation}
\label{eigenvs+}
F\leq F_{comp}(\mu):=\frac{d-\mu}{d^2(1-\mu)}\ .
\end{equation}
Then, the volume $V_{comp}(\mu,d)$ of entangled $\Lambda_\mu$-compatible isotropic states is the length of the interval $[1/d,F_{comp}(\mu)]$; namely,
\begin{equation}
\label{eigenvs++}
V_{comp}(\mu,d)=F_{comp}(\mu)-\frac{1}{d}=\frac{\mu(d-1)}{d^2(1-\mu)}\ .
\end{equation}
This volume vanishes as $d^{-1}$  for large $d$ isotropic states; namely, the larger $d$, the smaller is the set of entangled $\Lambda_\mu$-compatible isotropic states.

\section{Conclusions and outlook}
 
Whether Complete Positivity should or not be a necessary requirement of any legitimate physical transformation of an open quantum system $S$ is under ongoing debate.
The advocates of its physical necessity privilege dynamics $\Lambda$ that, when lifted to $\Lambda\otimes{\rm id}_n$, freely and consistently 
act on all possible, especially entangled, states of the compound system $S+S_n$ for any appended inert $n$-level system $S_n$.

Instead, to avoid unphysical effects as the appearance at some stage of negative probabilities, the dismissal of  Complete Positivity as a physically
unnecessary technical constraint forces one to restrict  the set of initial states of $S+S_n$ to those compound states that are compatible with $\Lambda\otimes{\rm id}_n$, that is to those  that are sent into states.

Regarding this way-out of Complete Positivity, we have shown that a problem arises in the specific context of isotropic states and for a specific family of channels; indeed, their compatible states 
become fewer and fewer with increasing dimension.
This is an indication of a possible intrinsic weakness that might be responsible for the depletion of initial states in the case of many-body open 
quantum dynamics that are described by non completely positive maps. 
Of course, it is important to assess how general is such a weakness, whether it affects more general states and dynamics than those considered in this 
manuscript. Such an assessment seems to be particularly significant in the case of the approach  invoking a so-called "slippage of initial conditions"~\cite{HartmannStrunz,Suarez,Gaspard} mechanism, namely the existence of a dynamical initial transient phase that provides the restriction to compatible states before the non-completely positive (or even non positive) open dynamics sets in.

\end{document}